\documentstyle[12pt]{article}

\begin{document}

\newcommand{\newc}{\newcommand}
\newc{\ra}{\rightarrow}
\newc{\lra}{\leftrightarrow}
\newc{\beq}{\begin{equation}}
\newc{\eeq}{\end{equation}}
\newc{\barr}{\begin{eqnarray}}
\newc{\earr}{\end{eqnarray}}
\def\ap{a^\prime}
\def\bp{b^\prime}
\def\al{\alpha}
\def\alp{\alpha^\prime}
\def\be{\beta}
\def\bep{\beta^\prime}
\def\gm{\gamma}
\def\dl{\delta}
\def\la{\lambda}
\def\ka{\kappa}
\def\gmp{\gamma^\prime}
\def\np{\nu^\prime}
\def\Gm{\Gamma}
\def\Sh{\hat S}
\def\mue{({\mu^-},{e^-})}
\def\nop{\nu_1^{\prime}}
\def\nwp{\nu_2^{\prime}}
\def\no{\nu_{1}}
\def\nw{\nu_2}

\def\fpt{F^{\prime2}}
\def\sF{sinF}
\def\sFt{{sin^2}F}
\def\cF{cosF}
\def\cf{cos{F\over2}}
\def\sf{sin{F\over2}}
\def\Wp{w^\prime}
\def\Wpt{w^{\prime2}}
\def\Ft{F_\pi^2}
\def\ft{{\Ft\over2}}
\def\gp{G^\prime}
\def\gpt{G^{\prime2}}
\def\go{\gamma_1}
\def\gw{\gamma_2}
\def\gt{\gamma_3}
\def\Wo{\Omega}
\def\wp{\omega^\prime}
\def\wpt{\omega^{\prime2}}
\def\bet{(\beta^\prime/2)}
\def\pa{\partial}
\def\ur{\underline r}
\def\ut{\underline \tau}
\def\bpp{\beta^{\prime\prime}}

\title{ ($\mu^-,e^-$) Conversion; A Symbiosis of Particle
and Nuclear Physics}

\author{T. S. KOSMAS  and  J. D. VERGADOS  \\
Theoretical Physics Division, University of Ioannina,
 GR 45110, Greece}
\date{}

\maketitle

\begin{abstract}
{ $\mue$ conversion is the experimentally most interesting lepton
flavor violating process. From a theoretical point of view it is an
interesting interplay of particle and nuclear physics. The effective
transition operator, depending on the gauge model, is in general
described in terms of a combination of four terms of transition
operators (isoscalar and isovector, Fermi-like as well as axial
vector-like). The experimentally most interesting ground state to ground
state transition is adequately described in terms of the usual proton
and neutron form factors. These were computed in both  the shell model
and RPA. Since it is of interest to know the portion of the strength
exhausted by the coherent (ground state to ground state) transition, the
total transition rate to all final states must also be computed. This
was done i) in RPA by explicitly summing over all final states ii) in
the context of the closure approximation (using shell model and RPA for
constructing the initial state) and iii) in the context of nuclear matter
mapped into nuclei via a local density approximation.

We found that, apart from small local oscillations, the conversion rate
keeps increasing from light to heavy nuclear elements. We also find that
the coherent mode is dominant (it exhausts more than 90\% of
the sum rule). Various gauge models are discussed. In general the
predicted branching ratio is much smaller compared to the present
experimental limit. }
\end{abstract}

\newpage
\section{ Introduction }

All currently known experimental data are consistent with the standard
model (SM) of weak and electromagnetic interactions.  Within the
framework of the SM, baryon and lepton quantum numbers are separately
conserved. In fact one can associate an additive lepton flavor quantum
number with each lepton generation which appears to be conserved. There
are thus three such conserved quantum numbers $L_e$, $L_{\mu}$ and $L_{\tau}$
 each  one associated with the lepton
generations $(e^-,\nu_e)$, $(\mu^-,\nu_{\mu})$,
$(\tau^-,\nu_\tau)$, with
 their antiparticles having opposite lepton flavor. It is
in fact these quantum numbers which distinguish between
the three neutrino species if they are massless.

Most theorists, however, view the SM not as the ultimate theory of
nature but as a successful low energy approximation.  In possible
extensions of the SM it is legitimate to ask whether lepton flavor
conservation still holds.  In fact in such gauge models (Grand
Unified Theories, Supersymmetric Extensions of the SM,
Superstring Inspired Models) such quantum numbers  are
associated with global (non local) symmetries and their
conservation must be broken at some level.

Motivated in part by this belief the search for lepton flavor
violation, which began almost  half a century ago (Hincks and
Pontecorvo, 1948 \cite{HING}, Lagarigue and Peyrou, 1952 \cite{LAGAR},
 Lokanathan and Steinberger, 1955 \cite{LOKA},
see also Frankel, 1975 \cite{FRAN}) has
been revived in recent years and is expected to continue in the near
future.  In the meantime the number of possible reactions for testing
lepton flavor has been increased.  The most prominent such reactions are

\beq
\mu \rightarrow e\gamma
\label{eq:1.1}
\eeq

\beq
\tau \rightarrow e\gamma \qquad and \qquad \tau \rightarrow \mu\gamma
\label{eq:1.2}
\eeq

\beq
\mu \rightarrow e e^+ e^-
\label{eq:1.3}
\eeq

\beq
\tau \rightarrow e e^+ e^- , \qquad    \tau \rightarrow \mu e^+ e^-
\label{eq:1.4}
\eeq

\beq
\tau \rightarrow e \mu^+ \mu^-, \qquad \tau \rightarrow \mu \mu^+\mu^-
\label{eq:1.5}
\eeq

\beq
K_L \ra \mu^{\pm} e^{\mp}, \qquad  K^+ \rightarrow \pi^+ \mu e
\label{eq:1.6}
\eeq

\beq
(\mu^+ e^-) \leftrightarrow (\mu^- e^+) \quad
 muonium-antimuonium \, oscillations
\label{eq:1.7}
\eeq

\beq
 \mu^- \, + \, (A,Z)\, \, \rightarrow \,\, e^- \, + \, (A,Z) \quad
 (muon-electron \, conversion)
\label{eq:1.8}
\eeq

Finally, one could have both lepton and lepton flavor violating
processes like

\beq
(A,Z) \rightarrow (A,Z{\pm}2) + e^{\mp} e^{\mp} \qquad
( \beta \beta_{o\nu} -decay)
\label{eq:1.9}
\eeq

\beq
 \mu^- \, + \, (A,Z) \rightarrow  e^+ \, + \, (A,Z-2) \qquad
(muon - positron\,\, conversion)
\label{eq:1.10}
 \eeq

{}From an experimental point of view the most interesting reactions
are (\ref{eq:1.1}), (\ref{eq:1.3}), (\ref{eq:1.8}), (\ref{eq:1.9})
and (\ref{eq:1.10}).
None of the above processes has yet been seen. The best limits obtained
are

\beq
R_{e\gamma} = \frac{\Gamma(\mu^+ \rightarrow e^+ \gamma)}
{\Gamma(\mu^+ \rightarrow e^+ \nu_e \bar{\nu}_\mu)} \,
< \,4.9\times 10^{-11} \qquad (90\% \,\,\, CL)
\label{eq:1.11}
 \eeq

\noindent
set by LAMPF (Bolton {\it et al.}, 1988 \cite{LAMPF}),

 \beq
R_{3e} = \frac{\Gamma(\mu^+ \rightarrow e^+e^-e^+ )
}{\Gamma(\mu^+ \rightarrow all)} \, < \, 1.0 \times10^{-12}
\qquad (90\% \,\,\, CL)
\label{eq:1.12}
 \eeq

\noindent
set by SINDRUM at PSI (Bellgardt {\it et al.,} 1988 \cite{BELL}),

\beq
R_{e^-N} \, = \, \frac{\Gamma(\mu^-Ti \rightarrow e^-Ti)}
{ \Gamma(\mu^- \rightarrow all)} \, <
 \, 4.6 \times10^{-12} \qquad (90\% \,\,\, CL)
\label{eq:1.13}
 \eeq

\noindent
set by TRIUMF (Ahmad {\it et al.,} 1987 \cite{AHM1}) using
a Time Projection Counter (TPC) and

\beq
R_{e^+N} = \frac{\Gamma(\mu^-Ti\rightarrow e^+Ca(gs))}
{ \Gamma(\mu^- \rightarrow all)}
\, < 4.4 \times10^{-12} \quad (90\% \,\, CL)
\label{eq:1.14}
 \eeq

\noindent
set by SINDRUM II (Badertscher {\it et al.}, 1991
\cite{BAD1}). For a $Pb$ target the corresponding $\mue$ conversion
limit is (Ahmad {\it et al.,} 1988 \cite{AHM1})

\beq
R_{e^-N} \, \, < 4.9 \times10^{-10} \qquad (90\% \,\,\, CL)
\label{eq:1.15}
 \eeq

{}From a theoretical physics point of view
the problem of lepton flavor non-conservation is connected with the
family
mixing in the leptonic  sector.  Almost in all models the above process
can proceed at the one loop level via the neutrino mixing.  However,
due to the GIM mechanism in the leptonic sector, the amplitude
vanishes in the limit in which the neutrinos are massless. In some
special cases the GIM mechanism may not be completely operative even
if one considers the part of the amplitude which is independent of
the neutrino mass (Langacker and London, 1988 \cite{GUTS}, Valle, 1991
 \cite{WAL},
Conzalez-Garcia and Valle, 1992 \cite{GWAL}). Even then, however, the process
is suppressed if the neutrinos are degenerate. It should be  mentioned that
processes (\ref{eq:1.1})-(\ref{eq:1.8}) cannot distinguish between Dirac
and Majorana neutrinos. Process (\ref{eq:1.9}) can
proceed only if the neutrinos are Majorana particles.

In more elaborate models one may encounter additional mechanisms for
lepton flavor violation. In Grand Unified Theories (GUT's) one may
have additional Higgs scalars which can serve as intermediate
particles at the one or two  loop level leading to processes
(\ref{eq:1.1})-(\ref{eq:1.8}).  In supersymmetric extentions
of the standard model one may encounter as intermediate particles the
superpartners of the above. Lepton flavor violation can also occur in
composite models, e.g. technicolor \cite{DIM}.  In fact,
 such models have already been ruled out by the present experimental
bounds (see eqs. (\ref{eq:1.11})-(\ref{eq:1.15})).

The observation of any of the processes
(\ref{eq:1.1})-(\ref{eq:1.10}) will definitely signal new physics
beyond the standard model.  It will severely restrict most models.  It
may take, however, even then much more experimental effort to unravel
specific mechanisms responsible for lepton flavor violation or fix the
parameters of the models.  The question of lepton flavor
non-conservation has been the subject of several review papers (Scheck,
1978 \cite{SCH}, Costa and Zwirner, 1986 \cite{COS}, Engfer and Walter,
1986 \cite{ENGF}, Vergados, 1986 \cite{VER},
Bilenky and Petcov, 1987 \cite{BILE},
Melese, 1989 \cite{MELE}, Heusch, 1990 \cite{HEN}, Herczeg, 1992 \cite{HER},
Schaaf, 1993 \cite{SCHA}, Kosmas, Leontaris and Vergados, 1994
\cite{KLVrev}). In the present
review we will focus our attention on recent theoretical developments of
the subject.  We will pay little attention to the experimental
situation, since we do not intend to dublicate the experimental
review which recently appeared (Schaaf, 1993 \cite{SCHA}).
Furthermore, the reader
can find an interesting account of the early experiments by Di Lella
\cite{LELL2}.

{}From a nuclear physics point of view the most interesting muon number
violating process is the $\mue$ conversion in eq. (\ref{eq:1.8}).
Its sister $(\mu^-,e^+)$
conversion in eq. (\ref{eq:1.10}) is much more complicated, since it involves
two nucleons. Furthermore, lepton violation is more likely to be
seen in neutrinoless double beta decay of eq. (\ref{eq:1.9}).
In this report we will focus our attention on the $\mue$
conversion.
We will concentrate mostly on the evaluation of nuclear matrix elements.
As we have mentioned above, experimentally the most important transition
is that to the ground state. It is, however, important to know which
fraction of the total strength goes to the ground state. We will,
therefore, also evaluate the total transition rate to all final states
in various nuclear models.

The $\mue$ conversion, compared to its competitors
$\mu \ra e \gamma$, $\mu \ra 3 e $ etc.,
has some experimental advantages \cite{KLVrev}:

1. The detection of only one particle is sufficient. No coincidence is
needed.

2. For electrons with the highest possible energy, i.e. $E_e \approx
m_{\mu} c^2 -\epsilon_b$ with $\epsilon_b$ the muon binding energy, the
reaction is almost background free. Indeed the reactions of induced
background are:

i) Muon decay in orbit. There is a tiny tail in the region of interest
which is proportional to $(E^{max}_{bg} - E_e)^5$, i.e. very small
($E^{max}_{bg}$ denotes the maximum energy of the background electrons).
But in this region the shape is known and it can be subtracted out.

ii) Radiative muon capture. Indeed this can be a source of background
since the photon can decay to $e^+e^-$ pairs as

\begin{eqnarray}
 \mu^- \, + \,(A,Z) \rightarrow (A,Z-1)\, + \, \nu_\mu  \, +
&{\gamma}&  \nonumber \\
&^{\mid}& \ra  e^+ \, +\, e^-
\label{eq:1.16}
\end{eqnarray}

If the neutrino and the positron carry away zero kinetic energy, the
background electron can be confused with interesting electron.
 We notice, however, that the maximum electron energy for process
 (\ref{eq:1.16}) is

\beq
E_{bg}^{max} = m_{\mu} c^2 -\epsilon_b -m_ec^2 -\Delta = E_e
- \Delta - m_e c^2
\label{eq:1.17}
\eeq

\noindent
where $\Delta$ is the difference in the binding energy of the two nuclei
involved in eq. (\ref{eq:1.16}). By a judicious choice of the target nucleus
$\Delta$ can be quite large (e.g. $\Delta =2.5 MeV$ for $^{12}C$).
Thus, one has a background free region if one restricts oneself to the
coherent mode (the final nucleus of eq. (\ref{eq:1.8}) is left in its ground
state).

\section{ Mechanisms for lepton flavor violation}

Obviously, in all models which allow $\mu \ra e\gamma$ to proceed
\cite{GUTS,KLVrev}, reactions of
eqs. (\ref{eq:1.8}), (\ref{eq:1.9}) and  (\ref{eq:1.10}) can also proceed
via a virtual photon (see fig. 1).
The dot in these figures indicates that the vertex is not elementary but
very complicated. Such processes are called photonic.
Muon-electron conversion can, of course, occur via mechanisms which do
not involve the photon (non-photonic mechanisms). These are Z-exchange
diagrams, i.e. the photon of fig. 1(b) is replaced by a Z-boson. In most
models these diagrams are less important than the photonic ones. Another
possibility is an effective 4-fermion contact interaction (box
diagrams).
We notice that here it is possible to have both protons and neutrons
participating which may lead to a nuclear enhancement (see fig. 2).

The most popular scenario for $\mue$ conversion involves intermediate
neutrinos, see fig. 3 for the photonic case and fig. 4 for the box-
diagram case. In these diagrams the neutrinos which propagate are the
mass eigenstates. The neutrinos involved in weak interactions,
$\nu_e$ and $\nu_{\mu}$, are not stationary states but linear
combinations of neutrino mass eigenstates given by $U_{\mu j}$ and
$U_{e j}$.
Both light $\nu_j$ and heavy $N_j$ intermediate neutrinos can propagate.
Since, however, the matrix $U$ is unitary, the part of the amplitude
which is independent of the neutrino mass vanishes (GIM mechanism).
Thus, the leading contribution is proportional to $\Delta
m_{\nu}^2/M_W^2$, for light neutrinos, or
$\Delta m_N^2 M_W^2/ m^4_N$, for heavy neutrinos \cite{VER}.
This means that the process is suppressed for neutrinos with masses very
different from the W-boson mass.
This is discouraging, since all reasonable models do not yield neutrinos
in the region of the W-mass.

Another possibility is to enlarge the Higgs sector. One can induce
$\mue$ conversion in models with two Higgs isodoublets. The most
dominant contribution, however, is expected to occur at the two loop
level \cite{KLVrev}. This calculation has only been done for the
$\mu \ra e \gamma$
reaction. Such calculations show that it is possible to reasonably
adjust the parameters of the model so that a branching ratio not far
from the experimental limit can be obtained. We should note, of course,
that for real photons one encounters only the electric dipole and
magnetic dipole form factors (see next section). For $\mue$ conversion
(virtual photons) one also needs the monopole form factors which have
not been calculated.
One expects, however, to be able to obtain large branching ratios when
this is accomplished for fast $\mu \ra e \gamma$ decay.

Other exotic scalars, like the singly charged $S^+$, the doubly charged
isosinglet $S^{++}$ and isotriplet $\chi^{++}$, can also lead to flavor
violations \cite{VER}. Such enlargements are not, however, favored by
recent
theoretical models, especially those which are superstring inspired, and
they are not going to be further discussed.

Another interesting possibility is the supersymmetric extension of the
standard model. One now has additional particles which are the
superpartners of the known particles.
With those extra particles participating as intermediaries one
can have a plethora of new diagrams \cite{KLVrev,KLV89}. The most
important for $\mue$ conversion are shown in fig. 5.

The mixing matrix $V$ entering the vertices is

\beq
V = S^+_e S_{\tilde e}
 \label{eq:1.18}
 \eeq

\noindent
where $S_e$ is the unitary mixing matrix for the charged leptons and
$S_{\tilde e}$ is the corresponding one for the S-leptons (see ref.
\cite{KLVrev}). To leading order (tree level) the matrices
$S_{e}$ and $S_{\tilde e}$ are the same, i.e. $V$ is diagonal, and
lepton flavor violation occurs. If, however, one goes beyond the tree
level and takes into account renormalization effects, the matrix $V$ is
no longer diagonal so that one can have lepton flavor violation
\cite{KLVrev}.

\section{ Expressions for the amplitude of $\mue$ conversion }

The amplitude for the $\mue$ conversion can be cast in the form
\cite{KLVrev}

\beq
{\cal M} =  \frac{4 \pi \alpha}{q^2}\,
J^{(1)}_{\lambda}  j^{\lambda}_{(1)}
+
\frac {\zeta}{m^2_{\mu}} \, J^{(2)}_{\lambda} j^{\lambda}_{(2)} \,
\label{eq:3.1}
\eeq

\noindent
where the first term is the photonic and the second
the non-photonic contribution.
$q$ is the momentum transfer and $\zeta$ takes the form

\beq
\zeta =
\cases{
\frac{G_Fm_{\mu}^2}{\sqrt 2}, & W-boson  \, mediated \, models\cr
m^2_{3/2}/m^2_{\tilde a}, & Supersymmetric \, models \cr }
\label{eq:3.2}
\eeq

\noindent
$m_{3/2}$ is the gravitino mass  and $m_{\tilde a}$ is the relevant
$s$-quark (supersymmetric partner of quark).

The hadronic currents are

\beq
J^{(1)}_{\lambda} = {\bar N} \gamma_{\lambda} \frac
{1 + \tau_3}{2} N,
\qquad (photonic)
\label{eq:3.3}
 \eeq

\beq
J_{\lambda}^{(2)} = {\bar N}\gamma_{\lambda} \frac{1}{2}\left[
(3 + \beta f_V\tau_3) + (f_V +f_A \beta \tau_3)\gamma_5 \right ] N,
\qquad (non - photonic)
\label{eq:3.4}
 \eeq

\noindent
($N = Nucleon$)
while the leptonic currents are

\beq
j^{\lambda}_{(1)} = {\bar u}(p_1) (f_{M1} + \gamma_5 f_{E1}) i
\sigma^{\lambda \nu} \frac{q_{\nu}}{m_{\mu}} +
 \left (f_{E0} + \gamma_5 f_{M0}\right) \gamma^{\nu}  \left (
 g_{\lambda \nu} - \frac {q^{\lambda} q^{\nu}}{q^2}\right )
\label{eq:3.5}
 \eeq

\beq
j^{\lambda}_{(2)} = {\bar u}(p_1)\gamma^{\lambda}
\, \frac {1}{2} \, ({\tilde f}_V +
{\tilde f}_A \gamma_5) u(p_{\mu})
 \label{eq:3.6}
 \eeq

\noindent
where $\beta =\beta_1/\beta_0$, is the ratio of the isovector to the
isoscalar component of the hadronic current at the quark
level. The form factors
$f_{M1},$ $f_{E1},$ $f_{E0},$ $f_{M0},$
${\tilde f}_V$ and ${\tilde f}_A $ as well as the parameter $\beta$
depend on the assumed gauge model and the mechanism adopted (the
isoscalar parameter $\beta_0$ is absorbed in the definition of
the leptonic form factors).
For purely left-handed theories the number of independent form factors
is reduced in half since

\beq
f_{E0} = -f_{M0}, \qquad
f_{E1} = -f_{M1}, \qquad
{\tilde f}_A  = - {\tilde f}_V
\label{eq:3.7}
\eeq

\noindent
In some models involving the $W$-boson one has

\beq
\frac {f_{E0}}{q^2} = - \frac{f_{E1}}{m^2}
\label{eq:3.8}
\eeq

\noindent
while in the supersymmetric models discussed above one finds

\begin{eqnarray}
f_{E0} &=& -f_{M0} = - \frac{1}{2}
\tilde{\eta} \alpha^2 g(x) \frac{m^2_{\mu}}{m^2_{3/2}} \\
4\pi \alpha f_{E1} &=& - 4\pi \alpha f_{M1} = - \frac{1}{2}
\tilde{\eta} \alpha^2 f(x) \frac{m^2_{\mu}}{m^2_{3/2}} \\
{\tilde f}_V &=& - {\tilde f}_A = - \frac{\beta_0}{2} \tilde{\eta}
\alpha^2 f_b(x) \frac{m^2_{\mu}}{m^2_{3/2}}
\label{eq:3.ar}
\end{eqnarray}

\noindent
and

\beq
\beta_0 = \frac{4}{9}+\frac{1}{9}\frac{m_{\tilde u}^2}{m_{\tilde d}^2}
,\quad
\beta_1 = \frac{4}{9}-\frac{1}{9}\frac{m_{\tilde u}^2}{m_{\tilde d}^2}
\label{eq:3.12}
\eeq

\noindent
The functions $g(x)$, $f(x)$, $f_b(x)$ depend on the quantity
$\gamma = m_{\tilde \gamma}/ m_{\tilde a}$.
Since, however, this quantity is much smaller than unity, we get

\beq
f(x) \, \approx \, \frac{1}{12}, \qquad
g(x) \approx \frac{1}{18}, \qquad
f_b(x) \approx \frac{1}{8}
\label{eq:3.13}
\eeq

\section{Effective nuclear transition operator and nuclear matrix
elements }

The first step in constructing the effective transition operator is to
take the non-relativistic limit of the hadronic currents
in eqs. (\ref{eq:3.3}), (\ref{eq:3.4}). This leads to
the operators \cite{KLVrev,KoVe,KVCF}

\begin{eqnarray}
\Wo_0 = \tilde g_V{\sum^{A}_{j=1}}\Big (
 {3+f_V\be\tau_{3j}}\Big)
e^{-i{\bf q} \cdot {\bf r}_j} ,\quad
{\bf \Wo} =-\tilde g_A {\sum^{A}_{j=1}}\Big (
{\xi +\be\tau_{3j}}\Big) {{{\bf \sigma}_j}\over
\sqrt 3}\,e^{-i{\bf q} \cdot {\bf r}_j}
\label{eq:4.1}
\end{eqnarray}

\noindent
with $\xi =f_V/f_A$ and

\beq
{\tilde g}_V=\frac{1}{6}, \quad {\tilde g}_A = 0, \quad f_V=1,
 \quad \beta =3 \quad (photonic \,\,\, case)
\label{eq:4.2}
\eeq

\beq
{\tilde g}_V={\tilde g}_A= \frac{1}{2},\quad f_V=1,\quad
f_A=1.24   \quad (non-photonic \,\,\, case)
\label{eq:4.3}
\eeq

\noindent
For neutrino mediated processes

\beq
\beta_0 = \cases{30 \cr 1 \cr}, \qquad
 \beta_1 = \cases{25 \cr 5/6  \cr} \quad \qquad
{light \,\, neutrinos \atop heavy \,\, neutrinos }
\label{eq:4.4}
\eeq

\bigskip
\noindent
i.e. $\beta =5/6 \, \approx 0.8$ in both cases. For the supersymmetric
model discussed above

\begin{equation}
\beta_0 \approx 5/9, \qquad  \beta_1 = \frac{1}{3}, \qquad i.e. \,\,
\,\, \beta = 0.6
\label{eq:4.5}
\end{equation}

\noindent
The factor $1/\sqrt{3}$ in ${\bf \Wo}$ was introduced for convenience.
Thus one has

\beq
\mid ME \mid ^2 \,\, =\,\,
f_V^2 \,\mid <f \mid \Omega_0 \mid i,\mu> \mid ^2
+3 f_A^2 \,\mid <f \mid {\bf \Omega} \mid i,\mu> \mid ^2
\label{eq:4.6}
\eeq

The second step is to factor out the muon 1s wave function
\cite{Prima,G-P} i.e.

\beq
\mid <f \vert \Omega \Phi_{\mu} ({\bf r}) \vert i > \mid ^2 \,
\approx \, < \Phi_{1s}^{2} > \mid <f \mid \Omega \mid i> \mid ^2
\label{eq:4.7}
\eeq

\noindent
where

\begin{equation}
< \Phi_{1s}^{2} > \equiv \frac{ \int d^{3}r \vert \Phi_{\mu} ({\bf r})
\vert^{2} {\cal \rho} ({\bf r})}{ \int d^{3} r {\cal \rho} ({\bf x})}
\label{eq:4.8}
\end{equation}

\noindent
with $\Phi_{\mu} ({\bf r})$ the muon wave function and ${\cal \rho}({\bf
r})$ the nuclear density.
The above approximation was recently found to underestimate the width in
heavy nuclei by as much as $40\%$ (see below sect. 5).
However, it is still a good approximation for the
branching ratio $R_{e^-N}$ of eq. (\ref{eq:1.13}),
since it affects both the
numerator and the denominator by the same way.

With the above operators one can easily proceed with the evaluation of
the relevant nuclear matrix elements. We distinquish  the following
two cases.

\subsection{The coherent $(\mu^{-}, e^{-})$ conversion matrix elements}

For $0^+\ra 0^+$ transitions only the operator $\Wo_0$ in eq.
(\ref{eq:4.6}) contributes. One finds \cite{KLVrev}

\begin{equation}
<i \mid \Omega_0 \mid i >\, = \tilde g_V \,(3 + f_V \beta)
 \,  F(q^2)
\label{eq:41.1}
\end{equation}

\noindent
where

\begin{equation}
F (q^2) = F_Z (q^2) + \frac{3-f_V\beta}{3+f_V\beta}
F_{N} (q^2)
\label{eq:41.2}
\end{equation}

\beq
{F}_{Z} (q^2) = \frac{1}{Z} \int d^{3}r {\cal \rho}_{p} ({\bf r})
e^{- i {\bf q} \cdot {\bf r}},
\quad
{F}_{N} (q^2) = \frac{1}{N} \int d^{3}r {\cal \rho}_{n} ({\bf r})
e^{- i {\bf q} \cdot {\bf r}}
\label{eq:41.3}
\eeq

\noindent
$F_Z$ and $F_N$ are the proton, neutron nuclear form factors
with  ${\cal \rho}_{p}({\bf r})$, ${\cal \rho}_{n}({\bf r})$ the
corresponding densities normalized to Z and N, respectively.

Then, the branching ratio $R_{e^- N}$ takes the form

\beq
R_{e^-N} = \frac {1} {(G_F m^2_{\mu})^2} \,  \{ \,\vert \frac{
m^2_{\mu}}{q^2}f_{M1}
+ f_{E0} + \frac {1}{2} \kappa {\tilde f}_V \vert ^2
+ \vert \frac{ m^2_{\mu}}{q^2}f_{E1} + f_{M0} +
\frac {1}{2} \kappa {\tilde f}_A \vert ^2 \, \}\,  \gamma_{ph}
\label{eq:41.4}
 \eeq

\noindent
where $\kappa$ and $\gamma_{ph}$ carry all the the dependence on the
nuclear physics, i.e.

\beq
\gamma_{ph} = \frac {Z |F_Z(q^2)|^2
}{G^2 f_{PR}(A,Z)}
\label{eq:41.5}
 \eeq

\noindent
and

\beq
\kappa =\left(1+\frac{N}{Z} \, \frac{3-\beta}{3+\beta} \,
\frac{F_N(q)^2}{F_Z(q)^2}\right) \zeta
\label {eq:41.6}
\eeq

\noindent
In eq. (\ref{eq:41.5}) $G^2$ is a combination of the coupling
constants entering the ordinary
muon capture $(G^2 \approx 6)$ and $f_{PR}$ is the well known Primakoff
function \cite{Prima,G-P} which adequately describes the
ordinary muon capture throughout the
periodic table. It is approximately given by
\cite{G-P}

\beq
f_{PR} \, \approx \, 1.6 \, \frac{Z}{A} - 0.62
\label{eq:41.7}
\eeq

It is sometimes convenient to factor out the nuclear dependence from the
dependence on the rest parameters of the theory
\cite{KV88}, i.e. we write

\beq
R_{e^-N} = \rho  \gamma
 \label {eq:41.8}
 \eeq

\noindent
where the quantity $\rho$ is independent on nuclear physics. The quantity
$\gamma$ takes the form

\beq
\gamma = \frac{ \mid ME \mid ^2 }{ G^2 Z f_{PR}(A,Z)}
 \label {eq:41.9}
 \eeq

\noindent
In the case of the supersymmetric model discussed above $\gamma$ takes
the form

\beq
\gamma \, = \, \left( \frac{1}{3}+
\frac{3\kappa}{4}\right)^2\gamma_{ph.}
\label{eq:41.10}
\eeq

At this point it is of interest to compare the branching ratio of $\mue$
conversion to that of the $\mu \ra e \gamma$ decay. One finds

\beq
\frac {R_{e^-N}}{R_{e\gamma}} \,  \approx \, \frac{\alpha}{6\pi}\left
(\frac{1}{3} + \frac{3\kappa}{4}\right)^2\gamma_{ph}
\label{eq:41.11}
\eeq

\subsection{ Total $\mue$ conversion matrix elements}

As it was mentioned in the introduction, only the coherent rate is of
experimental interest. It is, however, important to know what portion of
the sum rule is exhausted by the coherent mode. We thus have to evaluate
the total matrix element

\beq
 M^2_{tot} =
f_V^2 \,\, \sum_{f} \Big({{q_f} \over {{m_{\mu}}} } \Big)^2
\,\mid <f \mid \Omega_0 \mid i > \mid ^2 +
3 f_A^2 \sum_{f} \Big({{q_f} \over {{m_{\mu}}} } \Big)^2
\,\mid <f \mid {\bf \Omega } \mid i > \mid ^2
\label{eq:42.1}
\eeq

\noindent
where

\beq
q_f \, = \,\, m_\mu -\epsilon_b -(E_f-E_{gs})
\label{eq:42.2}
\eeq

\noindent
$E_f$, $E_{gs}$ are the energies of the final and ground state of the
nucleus. This evaluation
clearly can be done in a model in which the final states can be
explicitly constructed. This is a formidable task, however, and only in
simple models can easily be done (e.g. RPA \cite{KVCF,KFSV}).

The other alternative is to use some approximation scheme. The first is
the so-called "closure approximation"
\cite{KoVe,KV89}. In this approximation one first
replaces the momentum $q_f$ by a suitable average, i.e. $q_f \ra k =
<q_f>$. Thus \cite{KLVrev},

\beq
\Omega_0 ({\bf q}_f)
\, \approx \, \sum_j \, \omega_0 (j)
e^{-i{\bf k} \cdot {\bf r}_j} \, \equiv
\Omega_0 ({\bf k})
\label{eq:42.3}
\eeq

\beq
{\bf \Omega} ({\bf q}_f)
\, \approx \, \sum_j \, {\bf \omega} (j)
e^{-i{\bf k} \cdot {\bf r}_j} \, \equiv
{\bf \Omega} ({\bf k})
\label{eq:42.4}
\eeq

\noindent
where

\beq
\omega_0 (j) \, =\, 3 + f_V \tau_{3j}, \qquad
{\bf \omega } (j) \, = \, (\xi + \beta \tau_{3j} ){\bf \sigma}/\sqrt{3}
\label{eq:42.5}
\eeq

Then, we write

\begin{eqnarray}
S_{A} &=& {\sum_f} \Big({ {q_f}\over {m_{\mu}}}\Big)^2
{\mid <f \mid {\bf \Omega} ({\bf q}_f) \mid i > \mid}^2
\nonumber \\
&\approx& \frac{k^2}{m^2_{\mu}}\, \sum_f <i \mid {\bf \Omega}^+(k)
\mid f> \cdot < f \mid {\bf \Omega} (k) \mid i >
\label{eq:52.6}
\end{eqnarray}

\noindent
or using closure over the final states we get

\beq
S_A
\, = \, \frac{k^2}{m^2_{\mu}}\,  <i \mid {\bf \Omega}^+(k) \cdot
{\bf \Omega} (k) \mid i > \, = \, \frac{k^2}{m^2_{\mu}}\, \Big\{
<i \mid {\bf \Omega}_{1b} \mid i > \, + \, <i \mid {\bf \Omega}_{2b}
\mid i >  \Big\}
\label{eq:42.7}
\eeq

\noindent
where

\beq
{\bf \Omega}_{1b} \, = \, \sum_j \, {\bf \omega}^+(j) \cdot {\bf
\omega}(j)
\label{eq:42.8}
\eeq

\beq
{\bf \Omega}_{2b} \, = \, \sum_{i \ne j} \, {\bf \omega}^+(i) \cdot
{\bf \omega}(j)
e^{-i{\bf k} \cdot ({\bf r}_i - {\bf r}_j)}
\label{eq:42.9}
\eeq

The computation of $S_V$ is analogous with

\beq
\Omega_{1b} \, = \, \sum_j \, \omega^+_0 (j) \omega_0(j)
\label{eq:42.10}
\eeq

\beq
\Omega_{2b} \, = \, \sum_{i \ne j }\, \omega^+_0 (i) \omega_0 (j)
e^{-i{\bf k} \cdot ({\bf r}_i - {\bf r}_j)}
\label{eq:42.11}
\eeq

We thus find

\beq
M^2_{tot} \,= \, f^2_V S_V + 3 f^2_A S_A
\label{eq:42.12}
\eeq

The real question is what is the appropriate value of k? In earlier
calculations \cite{KoVe,KV89} an average excitation
energy $\bar E = 20 MeV$ was used which was
derived from the ordinary muon capture phenomenology. It was recently
realized, however, that the correct excitation energy must be appreciably
smaller \cite{KVCF}
due to the presence in $\mue$ conversion of the coherent
mode with $E_f =0$. This channel is absent in muon capture. It seems
reasonable,
therefore, to use an average energy defined by

\beq
 {\bar E} =\,\, { { {\sum_{f}} (E_f - E_{gs})
\, \Big({{\mid {\bf q}_f \mid} \over {m_{\mu}} }\Big)^2\, {\mid <f
\mid \Omega \mid gs > \mid}^2 } \over { {\sum_f}
\, \Big({{\mid {\bf q}_f \mid} \over {m_{\mu}} }\Big)^2\,
{\mid <f \mid \Omega \mid gs > \mid}^2 }}
\label{eq:42.13}
\eeq

\noindent
This definition cannot in practice be used since presupposes knowlendge
of the final states, which we like to avoid. It can, therefore, be
estimated in a simple model like RPA (see next section).
In the context of RPA we can also
estimate easily the effect of the $gs \ra gs $ correlations
by using the Thouless theorem and defining
a correlated vacuum in terms of the uncorrelated one (see ref.
\cite{KLVrev,KVCF}). This shows that both the coherent as well as the total
$\mue$ conversion matrix elements are a rescaling of the uncorrelated
ones and that the ground state correlations strongly reduce all matrix
elements (see sect. 5).

Another method for calculating the $\mue$ conversion widths
has recently been proposed by the Valencia group
\cite{Oset1,Oset2,Oset3}. This
method employs nuclear matter mapped into nuclei via a local density
approximation (L.D.A.) utilizing the relativistic Lindhard function.
It has been proved that this method reproduces very well the ordinary
muon capture data. In this process only the incoherent channels are
open. The method, however, has been also applied to the incoherent part
of $\mue$ conversion \cite{Chiang}.
The Lindhard function was computed taking into account particle-hole
excitations of p-n type in a local Fermi sea without invoking the
approximation of eq. (\ref{eq:4.7}). The coherent mode is evaluated
independently
by using a local density approximation in eq. (\ref{eq:4.6}).

\section{ Results and discussion }

As we have mentioned in the introduction, in the present report we focus
on the nuclear dependence of $\mue$
conversion rates. In this section we present and discuss the nuclear
matrix elements for both the ground state to ground state
transition (coherent mode) and the sum over all final states (total rate)
obtained in the context of the three methods discussed above, i.e. (i)
shell model, (ii) quasi-particle RPA and (iii) nuclear matter mapped
into nuclei.

\subsection{ Coherent $\mue$ conversion }

For the coherent mode one essentially needs only calculate the proron and
neutron nuclear form factors. In table 1 the results for the
$0^+ \ra 0^+$ transition matrix elements obtained in the framework of
the shell model and quasi-particle RPA are shown. They describe the
mechanism for the photonic diagrams ($\beta =3$) and non-photonic
ones ($\beta =5/6$) discussed in sect. 2. We see that the two methods
give about the same results.
In both methods we have followed the conventional approach and ignored
the muon binding energy (see table 1 results labeled shell model and QRPA(i)).
To check the validity of this approximation the calculation was repeated
in the context of RPA (see table 1 results labeled QRPA(ii))
by explicitly calculating the binding energy $\epsilon_b$.
For heavy nuclei
it was found that the consideration of $\epsilon_b$ gives
about $40\%$ larger results. This is expected, since in this case the
momentum transfer at which the form factors are calculated is smaller.

We mention that similar results have also been found by using
$\epsilon_b$ with the method of nuclear matter mapped into nuclei
\cite{Chiang}. These results are shown in table 2.
With the latter method the accuracy of using the
approximation of eq. (\ref{eq:4.7}) was also tested by doing
exact calculation of the muon-nucleus overlap which is possible in the
context of the nuclear matter mapped into nuclei method \cite{Chiang}.
We see that the exact calculations for the absolute rates
(case II of table 2) differ appreciably from
those of the approximation of eq. (\ref{eq:4.7}) (case I of table 2).
We should mention, however, that we expect this approximation to be much
better in the branching ratio provided that the total $(\mu^-,\nu_{\mu})$
rate is calculated in the same way. One then can use the
Primakoff's function (see eqs. (\ref{eq:41.5}) and (\ref{eq:41.7})) which
was derived under this approximation.

The effect of the ground state to ground state correlations was estimated
in the context of RPA as was discussed in sect. 4.2.
The results obtained for $^{48}Ti$ nucleus are shown in tables
3 and 4. We see that these correlations reduce both the coherent and
total matrix elements by about 30\% which is in agreement with other
similar results \cite{PElli,McNe}.

\subsection{ Incoherent $\mue$ conversion }

The incoherent $\mue$ conversion involves the matrix elements of all the
excited states of the participating nucleus. These have been calculated
with two methods: (i) by explicitly constructing the final states in the
context of the quasi-particle RPA \cite{KVCF} (see tables 3 and 4 results
labeled QRPA(explicit)) and (ii) by summing over
a continuum of excited states in a local Fermi sea in the framework of
the nuclear matter mapped into nuclei method \cite{Chiang}
(see results labeled L.D.A. in table 3).

The total rates have also been obtained by
assuming closure approximation and employing a mean excitation energy of
the studied nucleus as was discussed in sect. 4.2.
The results obtained this way both in the context of
shell model \cite{KoVe} and RPA sum rules \cite{KVCF}
are shown in tables 3 and 4 for various mean excitation energies
${\bar E}$ (see sect. 4.2 for details).
For RPA the mean excitation energy used
was found by firstly calculating one by one the excited states included
in eq. (\ref{eq:42.1}).
Since such a calculation in the context of the
shell model is very tedious, the value of $\bar E$ used in shell model
sum rules was chosen in analogy to that of the muon capture reaction.
The big difference appeared between the two values is justified because,
in the $\mue$ conversion the dominant $gs \ra gs$ transition (not
existing in the muon capture) contributes only in the non energy
weighted sum rule of eq. (\ref{eq:42.13}).
The above large mean excitation energy used in shell model sum rules
overestimated the incoherent matrix elements (especially in the region
of heavy nuclei), since they depend strongly on $\bar E$.

As we have stressed above, among all the open channels for $\mue$
conversion the transition to the ground state is of experimental
interest. A useful quantity is the ratio of the coherent rate divided by
the total $\mue$ rate, i.e.

\beq
\eta \, = \,
\frac{\mid ME \mid ^2_{coh}} {\mid ME \mid ^2_{tot}}
\label{eq:52.1}
\eeq

\noindent
By dividing the results obtained for the coherent mode
$(M^2_{gs \ra gs})$ and the total $\mue$
matrix elements $(M^2_{tot})$, either they are given directly from a sum
rule or by adding independent coherent and incoherent results, we found
that the coherent channel dominates (see tables 3 and 4) and that
throughout the periodic table $\eta \ge 90 \%$ (see ref.
\cite{KFSV} and \cite{Chiang}).

\section{ Summary  and conclusions }

In this report we have investigated the $(A,Z)\mue(A,Z)^*$ reaction
emphasizing its
dependence on nuclear physics. Using reasonable assumptions it was
possible to factor out the nuclear dependence. The nuclear matrix
elements of great interest were those entering the ground state to
ground
state transition. These were described in terms of proton and neutron
form factors. Shell model calculations give similar results with RPA
\cite{KVCF}.
The predicted proton form factors agree with those extracted
from electron scattering.
As expected \cite{PElli,McNe}
ground state correlations reduce the rates by as much as 30\%.
The predicted proton form factors agree with those extracted
from electron scattering.

The total matrix element was also computed. First, by explicit
calculations involving all the final states in the context of RPA.
Second, by employing a sum rule with a suitable mean excitation energy
$\bar E \approx 2 MeV$, i.e. much smaller than previously
expected \cite{KoVe}. This was done
both in shell model and RPA. Again the ground state correlations tend to
decrease the predicted rates. For the calculation of the total rate we
also used a new method which employs nuclear matter mapped into nuclei
via a local density approximation. Needless to say that, though all
three methods basically agree with each other, the accuracy of such
calculations is not as good as those for the coherent process. It is,
however, pretty certain to expect that throughout the periodic table the
coherent mode dominates, i.e.
$\eta \, \ge \, 90  \%$
This, as we have mentioned in the introduction, is experimentally very
important.

Returning to the coherent production we find that there is some
dependence on nuclear physics. In fact we find that

 \beq
1.67 \, \le \kappa \le 1.89
\label{eq:6.1}
\eeq

\noindent
(see eq. (\ref{eq:41.6})) which is a small effect.
The variation in $\gamma_{ph}$ is, however,
much more pronounced.

 \beq
1.6 \, \le \gamma_{ph} \le 26
\label{eq:6.2}
\eeq

\noindent
The $\mue$ conversion rate does not show a maximum around $A \approx
60$, as it was previously believed \cite{WeiFei}, but keeps increasing
all the way up to the heaviest elements \cite{Chiang}.

The nuclear dependence on $(A,Z)$ is not strong enough to overcome
the extra power of $\alpha$ in eq. (\ref{eq:41.11}). We find

 \beq
1.5 \times 10^{-3} \, \le \frac{R_{e^-N}}{R_{e\gamma}} \le
3 \times 10^{-2}
\label{eq:6.3}
\eeq

\noindent
(the largest refers to $^{132}Sn$ and the lowest to $^{12}C$).

Finally, it is not encouraging that the predicted branching ratios for
the muon number violating processes are much smaller than experiment.
In the supersymmetric model described above we find

\beq
1.2 \times 10^{-18} \, \le R_{e^-N} \le
2.4 \times 10^{-16}
\label{eq:6.4}
\eeq

\noindent
We must stress, however, that this should not discourage the
experimentalists, since we do not as yet have a complete theory to
adequately describe such processes.


\newpage
\begin{table}
\caption{ Coherent $\mue$ conversion matrix elements
$M^2_{gs \ra gs}$ calculated
in the context of shell model and quasi-particle RPA (see text).}
\begin{tabular}{lcccccc}
\hline
& & & & & & \\
Nucleus  & \multicolumn{3}{c}{ Photonic Mechanism ($\beta =3$) }
     & \multicolumn{3}{c}{ Non-Photonic Mechanism ($\beta = 5/6$)}\\
\hline
& & & & & & \\
$(A,Z)$&
Shell Model & QRPA (i) & QRPA (ii) &
Shell Model & QRPA (i) & QRPA (ii) \\
\hline
& & & & & & \\
$^{48}_{22}Ti_{26}$ & 142.7 & 135.2 & 139.6 & 374.3 & 363.2 & 375.2 \\
& & & & & & \\
$^{60}_{28}Ni_{32}$ & 187.5 & 187.8 & 198.7 & 499.6 & 498.2 & 527.4 \\
& & & & & & \\
$^{72}_{32}Ge_{40}$ & 212.9 & 212.7 & 227.8 & 595.8 & 596.2 & 639.5 \\
& & & & & & \\
$^{112}_{48}Cd_{64}$& 274.2 & 280.0 & 346.7 & 769.4 & 785.8 & 983.3 \\
& & & & & & \\
$^{162}_{70}Yb_{92}$& 313.6 & 311.0 & 484.3 & 796.0 & 840.3 &1412.1 \\
& & & & & & \\
$^{208}_{82}Pb_{126}$& 240.2 & 287.5 & 582.9 & 631.4 & 767.5 &1674.9 \\
\hline
\end{tabular}
\end{table}

\begin{table}
\caption{Coherent widths (in arbitrary units) for the photonic
and non-photonic mechanisms obtained with the exact muon wave function
(column labeled I) and with the approximation of eq.
(38) (column labeled II).}
\begin{tabular}{rrcccc}
\hline
  &    &                      &       &  & \\
   \multicolumn{2}{c}{Nucleus} &
   \multicolumn{2}{c}{non-photonic mechanism ($\beta=5/6$)} &
\multicolumn{2}{c}{photonic mechanism ($\beta=3$)}\\
\hline
  &    &                             &  &  & \\
A & Z &  I  &   II   &  I & II \\
\hline
  &   &   &    &                &       \\
12 &6 & 0.52 $10^{-4}$& 0.51 $10^{-4}$ & 0.21 $10^{-4}$
& 0.21 $10^{-4}$ \\
24 &12& 0.91 $10^{-3}$& 0.90 $10^{-3}$ & 0.38 $10^{-3}$
& 0.37 $10^{-3}$ \\
27 &13& 0.14 $10^{-2}$& 0.14 $10^{-2}$ & 0.56 $10^{-3}$
& 0.55 $10^{-3}$ \\
32 &16& 0.29 $10^{-2}$& 0.28 $10^{-2}$ & 0.12 $10^{-2}$
& 0.12 $10^{-2}$ \\
40 &20& 0.64 $10^{-2}$& 0.62 $10^{-2}$ & 0.27 $10^{-2}$
& 0.26 $10^{-2}$ \\
44 &20& 0.72 $10^{-2}$& 0.69 $10^{-2}$ & 0.26 $10^{-2}$
& 0.25 $10^{-2}$ \\
48 &22& 0.94 $10^{-2}$& 0.91 $10^{-2}$ & 0.35 $10^{-2}$
& 0.34 $10^{-2}$ \\
63 &29& 0.21 $10^{-1}$& 0.19 $10^{-1}$ & 0.78 $10^{-2}$
& 0.72 $10^{-2}$ \\
90 &40& 0.47 $10^{-1}$& 0.42 $10^{-1}$ & 0.17 $10^{-1}$
& 0.16 $10^{-1}$ \\
112&48& 0.62 $10^{-1}$& 0.54 $10^{-1}$ & 0.22 $10^{-1}$
& 0.19 $10^{-1}$ \\
208&82& 0.13 $10^0$   & 0.89 $10^{-1}$ & 0.41 $10^{-1}$
& 0.29 $10^{-1}$ \\
238&92& 0.12 $10^0$   & 0.73 $10^{-1}$ & 0.35 $10^{-1}$
& 0.22 $10^{-1}$ \\
\hline
\end{tabular}
\end{table}
\newpage

\begin{table}
\caption{Total and coherent $\mue$ matrix elements for $^{48}Ti$
(non-photonic mechanism, $\beta=5/6$). The ratio $\eta$ of eq.
(64) is also shown.}
\begin{tabular}{lrrcc}
\hline
             &           &                 &             &  \\
 $Method$ & $M_{gs \rightarrow gs}^2 $ & ${\bar E}$ & $M_{tot}^2 $ &
 $\eta$ ($ \% $) \\
 \hline
              &           &                 &             &  \\
 $L.D.A.$                  & 375.7 &       & 410.5 & 91.5  \\
 $Shell\,Model (sum-rule)$ & 374.3 & 20.0  & 468.0 & 80.0  \\
 $QRPA \, (explicit)     $ & 363.0 & -     & 386.4 & 93.9  \\
 $QRPA (sum-rule)$         & 363.0 & 0.5   & 366.2 & 99.1  \\
 $QRPA (sum-rule)$         & 363.0 & 5.0   & 376.5 & 96.4  \\
 $QRPA (sum-rule)$         & 363.0 & 20.0  & 382.8 & 94.8  \\
 $QRPA+Corr \,(sum-rule)$  & 236.2 & 0.5   & 238.6 & 99.0  \\
 $QRPA+Corr \,(sum-rule)$  & 236.2 & 5.0   & 245.4 & 96.3  \\
 $QRPA+Corr \,(sum-rule)$  & 236.2 & 20.0  & 249.5 & 94.7  \\
              &           &                 &             &  \\
\hline
\end{tabular}
\end{table}

\begin{table}
\caption{Total and coherent $\mue$ matrix elements for
the photonic mechanism ($\beta=3$). See caption of table 3. }
\begin{tabular}{lrrrc}
\hline
             &           &                 &             &  \\
 $Method$    & $M_{gs \rightarrow gs}^2 $ & ${\bar E}$  &
 $M_{tot}^2$ & $\eta$ ($ \% $) \\
\hline
             &           &                 &             &  \\
 $Shell\,Model (sum-rule)$  & 144.6 & 20.0  & 188.8 & 67.2  \\
 $QRPA \,(explicit)$        & 135.0 &  -    & 146.2 & 92.3  \\
 $QRPA  (sum-rule)$         & 135.0 &  1.7  & 138.3 & 97.6  \\
 $QRPA (sum-rule)$          & 135.0 &  5.0  & 140.6 & 96.0  \\
 $QRPA (sum-rule) $         & 135.0 & 20.0  & 141.7 & 95.3  \\
 $QRPA + Corr \,(sum-rule)$ & 87.8  &  1.7  & 90.4  & 97.1  \\
 $QRPA + Corr \,(sum-rule)$ & 87.8  &  5.0  & 91.8  & 95.6  \\
 $QRPA + Corr \,(sum-rule)$ & 87.8  & 20.0  & 92.6  & 94.8  \\
              &           &                 &             &  \\
\hline
\end{tabular}
\end{table}

\newpage
\centerline{ \bf Figure Captions}

\vspace{10mm}
\centerline{ \bf Fig.  1}

\bigskip
{ Photonic diagrams for the elementary process
$\mu \ra e\gamma$, fig. 1(a), and the induced processes
$\mue$ conversion, fig. 1(b), $\mu \ra e^+ e^-$, fig. 1(c) and
muonium-antimuonium oscillations, fig. 1(d). $\bullet$ is a complicated
vertex. }

\vspace{10mm}
\centerline{ \bf Fig.  2}

\bigskip
{ Box-diagrams of $\mue$ conversion in an effective 4-fermion
contact interaction. Both protons and neutrons participate.}

\vspace{10mm}
\centerline{ \bf Fig.  3}

\bigskip
{ Photonic diagrams of $\mue$ conversion in a model with mixed
intermediate neutrinos.}

\vspace{10mm}
\centerline{ \bf Fig.  4}

\bigskip
{ Box diagrams (for protons and neutrons)
of $\mue$ conversion in a model with mixed intermediate
neutrinos. }

\vspace{10mm}
\centerline{ \bf Fig.  5}

\bigskip
{ SUSY diagrams for $\mue$ conversion.
Intermediate charged s-lepton mixing. }

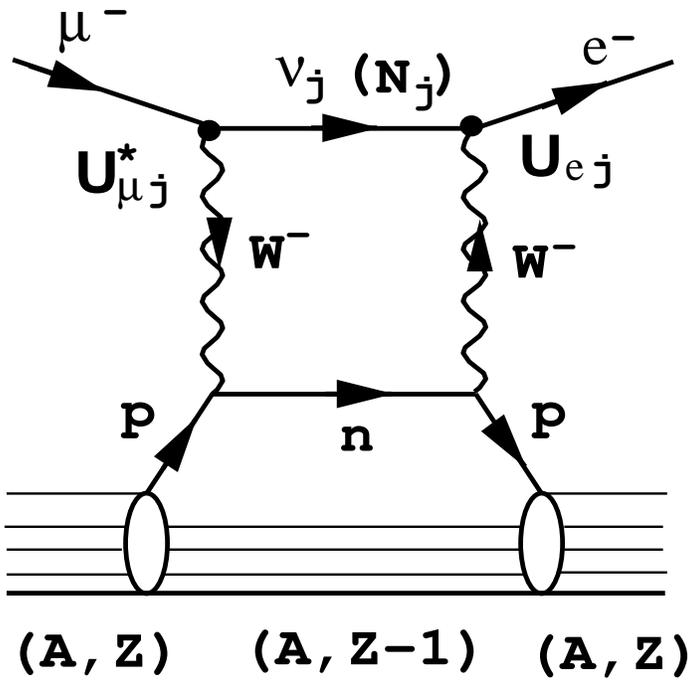
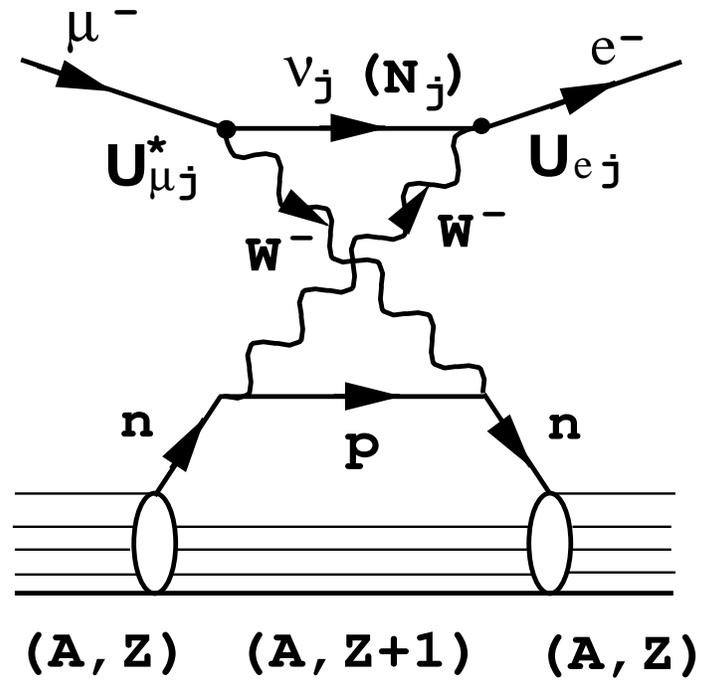
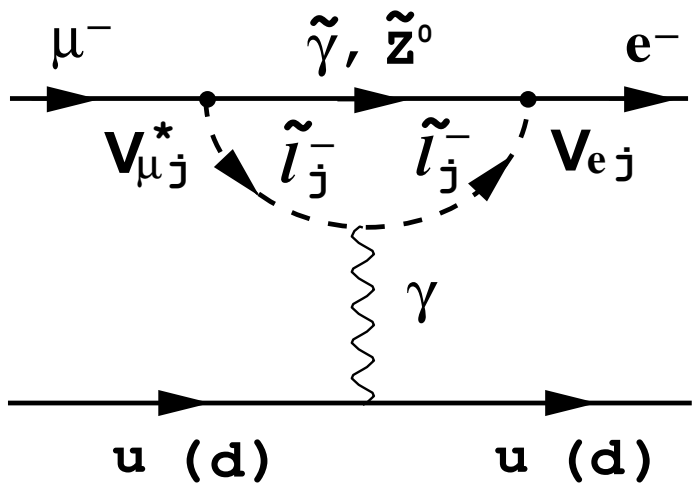
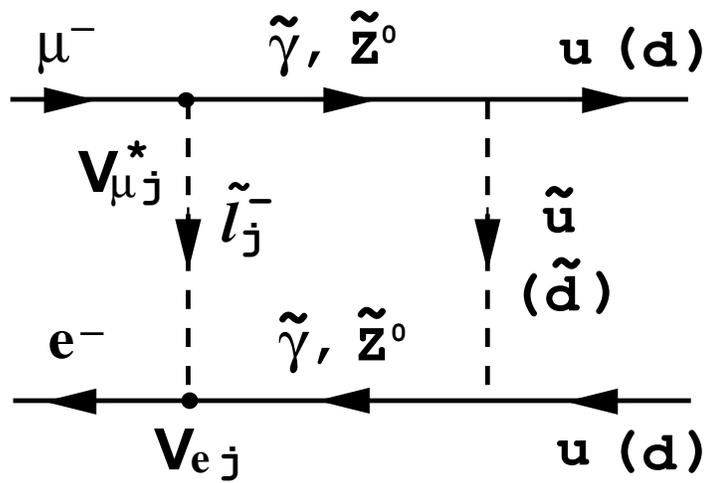
\end{document}